\documentclass[conference]{IEEEtran}
\IEEEoverridecommandlockouts

\usepackage{cite}
\usepackage{url}
\usepackage{amsmath,amssymb,amsfonts}
\usepackage{algorithmic}
\usepackage{multirow}
\usepackage{graphicx}
\usepackage{textcomp}
\usepackage{xcolor}

\def\BibTeX{{\rm B\kern-.05em{\sc i\kern-.025em b}\kern-.08em
    T\kern-.1667em\lower.7ex\hbox{E}\kern-.125emX}}
\begin{document}

\title{Edge-based Denoising Image Compression}

\author{
\IEEEauthorblockN{  Ryugo Morita\IEEEauthorrefmark{1}\IEEEauthorrefmark{3}, 
                    Hitoshi Nishimura\IEEEauthorrefmark{2}, 
                    Ko Watanabe\IEEEauthorrefmark{3},
                    Andreas Dengel\IEEEauthorrefmark{3},
                    Jinjia Zhou\IEEEauthorrefmark{1}}
\IEEEauthorblockA{\IEEEauthorrefmark{1}Faculty of Science and Engineering, Hosei University, Tokyo, Japan}
\IEEEauthorblockA{\IEEEauthorrefmark{2}KDDI Research, Inc., Saitama, Japan}
\IEEEauthorblockA{\IEEEauthorrefmark{3}RPTU Kaiserslautern-Landau {\&} DFKI GmbH, Kaiserslautern, Germany}
}
\maketitle

\begin{abstract}
In recent years, deep learning-based image compression, particularly through generative models, has emerged as a pivotal area of research.
Despite significant advancements, challenges such as diminished sharpness and quality in reconstructed images, learning inefficiencies due to mode collapse, and data loss during transmission persist. 
To address these issues, we propose a novel compression model that incorporates a denoising step with diffusion models, significantly enhancing image reconstruction fidelity by sub-information(e.g., edge and depth) from leveraging latent space.
Empirical experiments demonstrate that our model achieves superior or comparable results in terms of image quality and compression efficiency when measured against the existing models.
Notably, our model excels in scenarios of partial image loss or excessive noise by introducing an edge estimation network to preserve the integrity of reconstructed images, offering a robust solution to the current limitations of image compression.
\end{abstract}


\section{Introduction}
Deep learning-based image compression systems have recently gained significant traction in the research area. 
These systems, as highlighted in studies\cite{agustsson2019generative} \cite{duan2023lossy} are increasingly challenging traditional engineered codecs like WebP, JPEG2000\cite{taubman2002jpeg2000}, and BPG, which is a leading engineered codec. This evolution signifies the growing efficacy and potential of deep learning techniques in the realm of image compression.
Particularly, generative models like Variational Autoencoders (VAEs) \cite{balle2018variational} and Generative Adversarial Networks (GANs) \cite{goodfellow2014generative} have garnered attention in this domain. These models are capable of reconstructing images from compressed data by reducing redundancy, while also maintaining a quality akin to the original.

However, existing generative model-based compression faces several challenges. (1) compression leads to less sharp reconstructed images, degrading the quality of the original image's restoration. (2) challenges such as mode collapse during the training phase can significantly hinder efficiency. (3) data loss during transmission necessitates re-sending images, leading to inefficiency and increased data usage. These issues are not unique to image compression but are also prevalent in text-to-image tasks, as noted in studies \cite{richardson2021encoding} \cite{harvey2021conditional}.

In response to these challenges, we introduce a new image compression model that leverages the strengths of diffusion models to significantly enhance the fidelity of image reconstructions. 
Specifically, we introduce the Edge Estimation Network to make our model obtain edge information from transmit latent which is encoded with VAEs.
The obtained edge information guides the denoising step of the diffusion model and solves the problem which is reconstructing the unsharp image. 
Edge information increases the overall image reconstruction capability by increasing the ability to reconstruct not only sharp images but also complex foreground images, which is due to image bias, the property that foreground images are often complex and background images are often simple. 
Additionally, our approach circumvents the common training hurdles by leveraging a foundational model pre-trained on a huge dataset, rather than relying on a narrowly specific dataset. 
This not only streamlines the training process but also enhances the model's robustness and generalizability.

Existing models have achieved high compression ratios.
But in a noisy environment, it comes to reconstructing images with incomplete information, often requiring a cumbersome trial-and-error process with image inpainting models after the reconstruction of the received image or re-sending images in the actual application. 
Our model leverages the robust generative capabilities of stable diffusion, enabling efficient image inpainting during transmission. This not only addresses the issue of data loss but also enhances the overall efficiency and effectiveness of the image compression process.
Through extensive qualitative and quantitative experimentation, our model has demonstrated the ability to match or even surpass the image compression capabilities of existing studies, marking a significant advancement in the field. 

\begin{figure*}[t]
    \centering
    \includegraphics[width=0.88\linewidth]{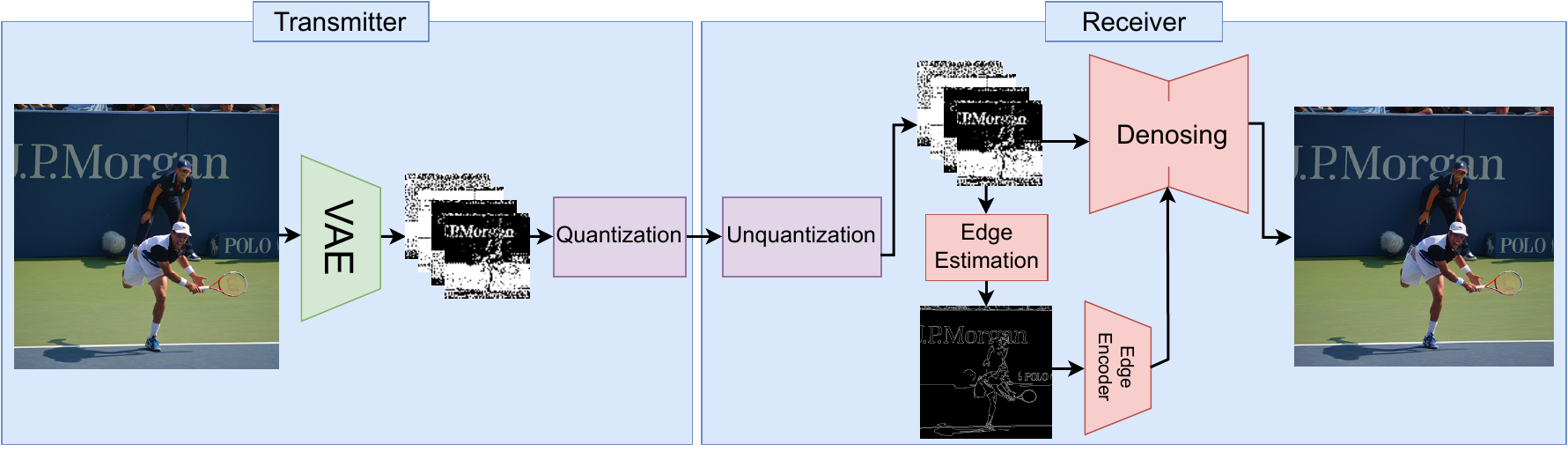}
    \caption{The overview of our proposed model.}
\label{model}
\end{figure*}

\section{Related Works}

\textbf{Image Compression} Classical codecs such as JPEG2000 \cite{taubman2002jpeg2000}, BPG, and WEBP have set longstanding benchmarks in image compression. However, they are now being increasingly challenged by end-to-end learned codecs \cite{wang2022neural} \cite{he2022elic}. These modern codecs utilize non-linear transform coding paradigms and are optimized for both rate and distortion metrics. Recently, there has been a growing interest in models that enhance realism in compression \cite{agustsson2019generative} \cite{mentzer2020high}. This trend has led to the optimization of autoencoder-based models with adversarial loss, though GAN training can be unstable. Despite these advancements, existing methods often suffer from reduced sharpness in reconstructed images and training inefficiencies. Our proposed model addresses these issues by incorporating edge information through an Edge Estimation Network and using a pre-trained foundational model, enhancing image sharpness and training stability.

\textbf{diffusion model} Recent advancements in image compression have been significantly influenced by probabilistic diffusion models, which have demonstrated exceptional performance in image generation tasks. These models often match or surpass the perceptual qualities of finely-tuned GANs while maintaining stable training processes \cite{ho2020denoising}\cite{yang2022diffusion}. Notable examples include Dall-E \cite{ramesh2022hierarchical} and Stable-Diffusion \cite{rombach2022high}.Research in this area includes \cite{hoogeboom2021autoregressive}, which explored an autoregressive diffusion model for lossless compression, and \cite{theis2022lossy}, which introduced a novel approach using unconditional diffusion models for lossy transmission. More recently, \cite{yang2024lossy} adapted the conditional diffusion model for image reconstruction. These studies represent a significant shift from traditional image compression techniques. However, these models often face challenges with data loss during transmission, leading to inefficiencies. Our model addresses this by leveraging stable diffusion's robust generative capabilities for efficient image inpainting during transmission, thus mitigating data loss and enhancing overall compression efficiency.

\section{Methods}
The architecture of our proposed model, as shown in Fig. \ref{model}, is designed to improve image compression efficiency. 
Our model initiates this process with an input image $I$, aiming to accurately reconstruct $I'$ from its latent representation.
Initially, the image undergoes encoding through a Variational Autoencoder (VAE) into a latent space $z$, utilizing advanced quantization techniques to facilitate efficient transmission. Upon reception, the latent space is unquantized to $z'$.
To address discrepancies between $z$ and $z'$ and improve the fidelity of the reconstructed image $I'$, we employ a pre-trained diffusion model for denoising. This model refines $z'$, mitigating the effects of quantization. However, to further combat the issue of blurred reconstructions, we integrate an edge estimates network. This network extracts crucial edge information $I_{edge}$ from $z'$, enabling conditional denoising that leverages both latent space and edge details for enhanced reconstruction quality.
By incorporating edge information in the denoising process, our model demonstrates remarkable resilience against noise and artifacts, reconstructing images with unparalleled accuracy even in challenging conditions. The subsequent sections will delve into the specifics of each architectural component, elucidating their roles and synergies in achieving superior image reconstruction.

\subsection{Edge Estimation Network(EEN)}
To enhance image sharpness, we introduce an Edge Estimation Network (EEN) into our framework. This innovative component is engineered to enhance the quality of reconstructed images by accurately detecting and incorporating edge information for the conditional denoising phase. Drawing inspiration from the work by \cite{morita2023batinet} and adapting techniques from naive image style transfer models \cite{isola2017image}, our EEN is specifically designed to process the unquantized latent representation $z'$, estimating the edge information $I_{edge}$. This is achieved by training the EEN generator to take the resized unquantized latent $z'$ as input and produce the edge information $I_{edge}$ as output, thereby allowing for the precise estimation of edges from the latent space. Further details on the training process and optimization strategies employed for the EEN will be discussed in Section \ref{sec:train}.

\begin{figure*}[t]
\centerline{\includegraphics[width=0.8\linewidth]{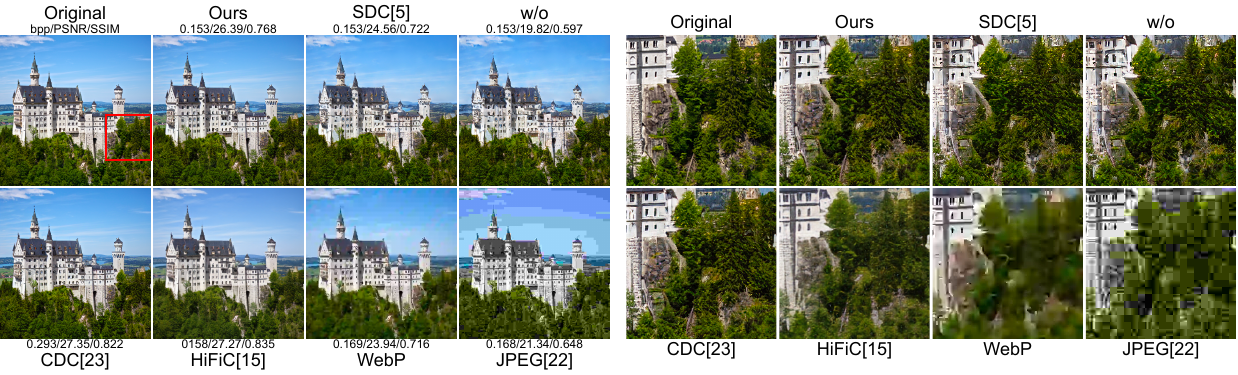}}
\caption{Qualitative comparison of the proposed method and the existing methods with smallest the bbp in each methods. Our model has a very high foreground image reconstruction system due to the acquisition of the edge information. It indicates the contours of the tree and castle are so sharp and refine compare with the existing model.}
\label{quantitative}
\end{figure*}

\begin{figure}[t]
\centerline{\includegraphics[width=0.8\linewidth]{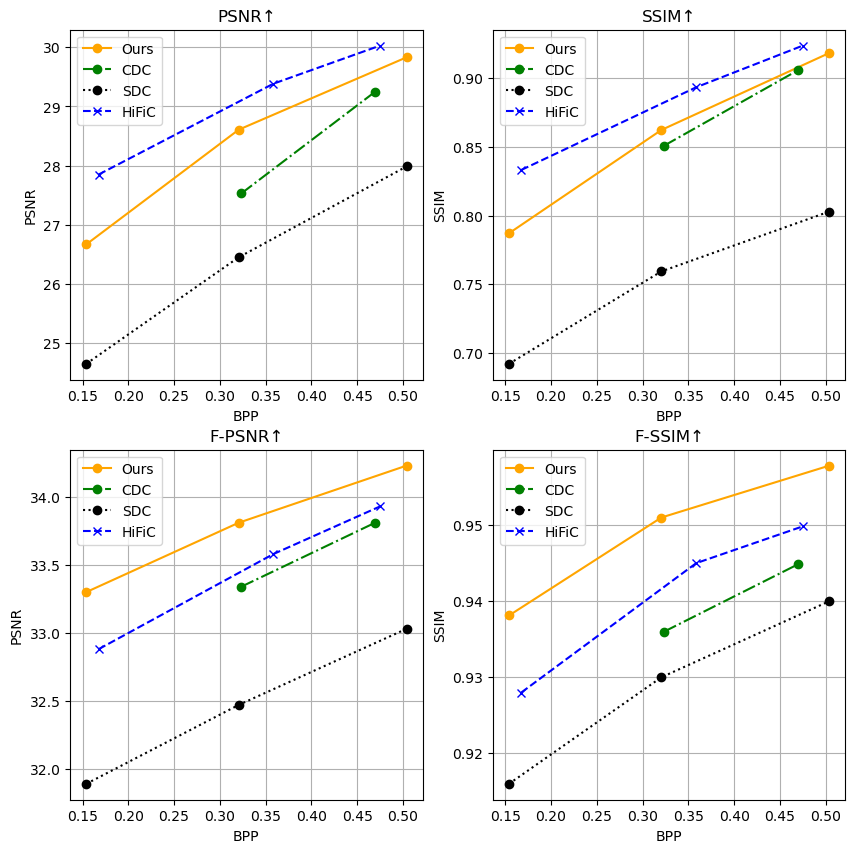}}
\caption{Quantitative result comparison of the proposed method and the existing methods on the DIV2L dataset. Our model outperforms the existing diffusion image compression model. 
F-PSNR and F-SSIM indicate the PSNR and SSIM for the foreground images. These images were masked by using a one-hot segmentation map detected by deeplabv3\cite{chen2018encoder}.}
\label{qualitative}
\end{figure}

\subsection{Canny-based Denosing Step(CDS)}
Our proposed model leverages the state-of-the-art text-to-image model, Stable Diffusion\cite{rombach2022high}, as a baseline. This model excels by projecting noisy images into latent space and effectively denoising them to reconstruct realistic images, even from noise-corrupted latent spaces. We utilize this ability to reconstruct realistic images from compressed latent spaces. However, Stable Diffusion alone cannot reconstruct the original image accurately due to its training on a dataset that differs from the original dataset. To address this, we adopt a conditional diffusion model, which is tailored to generate images based on specific conditions like open-pose, canny edge, and depth map. 
Inspired by ControlNET \cite{zhang2023adding}, a conditional diffusion model designed for text-to-image generation tasks, our model is enabled to more accurately adhere to the nuances of the original image while handling diverse image generation scenarios. Further details on the training process and optimization strategies employed for the CDS will be discussed in Section \ref{sec:train}.

\subsection{Temporary Data Complement Step}
In existing research, if data were lost during transmission, there is no way to deal with it while processing. 
In this case, the existing studies require resending or performing image processing, such as image inpainting without a reference image, after receiving. 
In contrast, our approach leverages the generative capabilities of diffusion models, which are particularly effective in situations where noise or data loss exceeds a certain threshold. It allows for the generation of complementary information at the receiver's end, thus minimizing transmission overhead.

For parts of the image identified as missing or corrupted, a mask is created by the sender. 
Accompanying this mask, edge information is derived from the Edge Estimation Network (EEN), which aids in the inpainting process. 
In scenarios where the image's integrity is compromised mainly by missing elements, we utilize the inpainting capabilities of ControlNET, directly applying its original weights for efficient restoration. 
Unlike the method of inpainting with text after transmission, this model considers the style of the original latent at the time of transmission, resulting in high inpainting accuracy without any text.
This method not only improves the accuracy of the reconstructed image but also substantially reduces the necessity for re-sending the entire image, offering a more efficient solution to overcoming transmission-related image degradation.

\renewcommand{\arraystretch}{1.2}
\begin{table*}[ht]
\caption{Comparison of Image Compression Performance for all and foreground image}
\centering
\begin{tabular}{|l|c|c|c|c|c|c|c|c|}
\hline
\multirow{2}{*}{} & \multicolumn{2}{c|}{SDC\cite{diffusioncompression2022}} & \multicolumn{2}{c|}{CDC\cite{theis2022lossy}} & \multicolumn{2}{c|}{HiFIC \cite{mentzer2020high}}& \multicolumn{2}{c|}{Ours} \\ \cline{2-9} 
& bpp & PSNR/SSIM & bpp & PSNR/SSIM & bpp & PSNR/SSIM & bpp & PSNR/SSIM 
\\ 
\hline
\hline
ALL        &  \multirow{2}{*}{0.320} & 26.45/0.759   & \multirow{2}{*}{0.323}  & 27.53/0.851   & \multirow{2}{*}{0.359} & \textbf{29.38}/\textbf{0.893}   & \multirow{2}{*}{0.320}  & 28.61/0.862
\\
Foreground &  & 32.47/0.930   &  & 33.34/0.936   &  & 33.58/0.945   &   & \textbf{33.81}/\textbf{0.951}
\\  
\hline
\end{tabular}

\label{tab:comparison}
\end{table*}

\subsection{Loss function and training}
\label{sec:train}
\textbf{Edge Estimation Network.}
Drawing from the framework proposed by \cite{isola2017image},  the Edge Estimation Network's generator loss, denoted as $L_{G_{een}}$, is formulated by combining an adversarial loss, $L_{adv}$, with an L1 loss, $L_{L1}$. The formulation is given as:
\begin{equation}
L_{L1} = \mathbb{E}{x,y,z} [|y - G(x,z)|]
\end{equation}
\begin{equation}
L{G_{een}} = L_{adv} + L_{L1}
\end{equation}
The discriminator's loss function is solely constituted by the adversarial loss, described by:
\begin{equation}
L_{D} = L_{adv}
\end{equation}
This structure ensures that while the discriminator focuses on distinguishing between real and generated edges, the generator is optimized to produce edges that are not only realistic but also closely aligned with the target, balancing the adversarial drive with the precision of L1 loss.

\textbf{Conditional Denoising step.}
In the training phase, our approach aligns with the methodology established by the original ControlNet, maintaining the generative model's weights while focusing updates on the canny edge encoder exclusively. This strategy facilitates the reconstruction of the original image without relying on textual cues. Utilizing a dataset comprising one million images from the Laion-Aesthetics dataset \cite{schuhmann2022laion}, we adhere to the objective function prescribed for standard diffusion models, which is expressed as follows:
\begin{equation}
L = \mathbb{E}{z_0, t, c_c, \epsilon \sim \mathcal{N}(0,1)} [|\epsilon - \epsilon\theta(z_t, t, c_c)|]
\end{equation}
Here, $L$ represents the training objective for the diffusion model, pivotal in the fine-tuning process for image reconstruction tasks. The variables $z_0$, $z_t$, $t$, and $c_c$ correspond to the original image, the image perturbed by noise, the specific timestep, and the conditions for control, respectively. This formulation underscores our tailored approach to refining the diffusion model's capacity for high-fidelity image reconstruction under conditional constraints.

\section{Experiment}
In our analysis, we engaged with three well-known datasets: Kodak \cite{kodak2013}, Tecnick \cite{asuni2014testimages}, and DIV2K \cite{agustsson2017ntire}, to validate our approach. The Kodak dataset is comprised of 24 high-resolution images (768×512 pixels), selected for their varied color schemes and compositions. The Tecnick dataset offers 100 images depicting natural scenes (originally 600×600 pixels), while the DIV2K dataset provides a validation set of 100 high-definition images of assorted dimensions. To maintain uniformity in our assessments, all images were resized and cropped to a uniform resolution of 512×512 pixels. This uniformity was essential for our evaluation, allowing for direct and reliable comparisons across different methods and reinforcing the credibility of our results.

In our investigation into image compression and the capability for interim image restoration, we adopted quantization strategies as detailed in the study by \cite{diffusioncompression2022}. Following their established quantization protocols, we aimed to evaluate our system's resilience to transmission errors by introducing scenarios of partial image loss. To this end, we implemented tests where random segments of an image, covering up to one-eighth of its dimensions, were obscured to determine the effectiveness of our reconstruction techniques using supplementary data. The recovery phase leveraged ControlNET’s inpainting model weights, facilitating an accurate and seamless restoration of the omitted sections of the images.

\subsection{Image compression Result}
\textbf{Qualitative comparison} Fig. \ref{qualitative} compares the existing models and ours on the Kodak dataset. 
It indicates that the output image lacks a sharp line with a diffusion-based model.
In contrast, our proposed model can generate sharper images because we adapt the EEM to utilize the edge information.
Furthermore, compared with the existing image compression model, ours outperforms reconstructions of the foreground content, such as a castle and the tree in the image.
This indicates that when the EES gets the edge information clearly, our model can focus on foreground image reconstruction to refine the image.

\textbf{Quantitative comparison} In the quantitative experiment, we evaluate the PSNR and SSIM on the DIV2L dataset, which is shown in Fig. \ref{quantitative}.
F-PSNR and F-SSIM indicate the PSNR and SSIM for the foreground images. These images were masked by using a one-hot segmentation map detected by deeplabv3 \cite{chen2018encoder}.
Ours outperforms existing diffusion-based models in all metrics. It indicates the edge information from the EEM makes the conditional denoising improve the accuracy of sharp image reconstruction.
On the other hand, our model does not exceed the accuracy of existing models, it maintains comparable or slightly lower performance levels. 
However, Fig. \ref{quantitative} in the second row and Table. \ref{tab:comparison} indicates that our model outperforms existing models in foreground images, even if the metrics of overall images are lower than ours.

Furthermore, ours utilizes a pre-trained model that is not tailored to any specific dataset, enabling robust image reconstruction across a variety of real-world images. 
This generalizability highlights the model's adaptability and efficiency, offering significant value for applications requiring reliable image processing in diverse conditions. Additionally, our model is capable of temporary data complement in compression, and the details of this task are described in Section \ref{sec:Temporary}.

\begin{figure}[t]
\centerline{\includegraphics[width=0.8\linewidth]{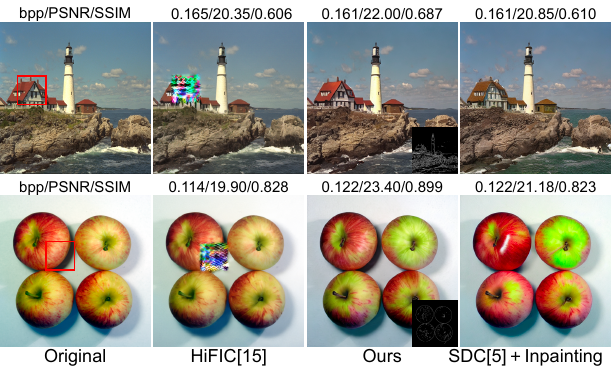}}
\caption{Temporary Data Complement of the proposed method and other existing methods when the vector or latent information misses the region while transmission. “Inpainting" indicates the inpainting with diffusion model after receiving the reconstructed image. To compare with ours, the inpainting cannot consider the latent and edge information to complete the image, in contrast, ours can do it. The existing model doesn't have this solution for completing the image in missed the information.}
\label{inpainit}
\end{figure}

\textbf{Ablation study} In Fig. \ref{quantitative} and Fig. \ref{qualitative}, SDC and W/O are the two models in this proposal: SDC is ours w/o EEN and conditional information, and W/O is ours w/o EEN, conditional information and denoising step.
Compared with the SDC, better results were obtained both quantitatively and qualitatively, because the input of conditional information solves the problem of reconstructing non-sharp images, which is a weak point of the VAE generation results used in this study.
Furthermore, a comparison with the W/O shows assistance for reconstruction in the denoising step.
This is because latent diffusion is essentially a task of generating images by denoising noise from latent and text information, so the noise due to some transmission is considered noise in the diffusion process.
Therefore, it was found that denoising the noise caused by some transmission can be considered as the noise in the diffusion process and the reconstruction error can be reduced by denoising the noise together.
The ablation study indicates that the proposed architectures were found to be robust to some extent even in the noise of image compression.

\subsection{Temporary Data Complement}
\label{sec:Temporary}
Fig. \ref{inpainit} showcases our model's unique ability to reconstruct images in scenarios where significant noise or data loss occurs in the latent space. 
Traditional image compression methods fail to reconstruct images if crucial information is missing during transmission. In contrast, our model successfully reconstructs images using edge and latent information, demonstrating resilience against data loss.
Through quantitative evaluations using the SSIM and PSNR, we've shown our model's superior reconstruction accuracy compared to prior studies. They mark a significant improvement over existing methods, which often require re-sending the image for minor data losses.
Our innovative "Temporary Data Complement" technique allows for image transmission without full data resend, addressing a previously unmet need in image compression applications and representing the first solution of its kind.

\section{Conclusion}
In conclusion, we introduces a novel image compression model that effectively addresses the prevalent challenges in generative-based model. By leveraging diffusion models for denoising training in the latent space, our approach significantly enhances the fidelity of image reconstruction. The empirical results demonstrate the superiority of our model over existing methods, particularly in image quality and compression efficiency. Additionally, the model's unique capability to handle scenarios of partial image loss or excessive noise by implementing temporary measures further underscores its robustness. This innovative feature not only improves performance but also broadens the practical applicability of our model in real-world settings, offering a comprehensive solution to the limitations of current image compression technologies.

\bibliographystyle{plain}
\bibliography{ref.bib}

\begin{thebibliography}{10}

\bibitem{agustsson2017ntire}
Eirikur Agustsson and Radu Timofte.
\newblock Ntire 2017 challenge on single image super-resolution: Dataset and
  study.
\newblock In {\em Proceedings of the IEEE conference on computer vision and
  pattern recognition workshops}, pages 126--135, 2017.

\bibitem{agustsson2019generative}
Eirikur Agustsson, Michael Tschannen, Fabian Mentzer, Radu Timofte, and Luc~Van
  Gool.
\newblock Generative adversarial networks for extreme learned image
  compression.
\newblock In {\em Proceedings of the IEEE/CVF International Conference on
  Computer Vision}, pages 221--231, 2019.

\bibitem{asuni2014testimages}
Nicola Asuni and Andrea Giachetti.
\newblock Testimages: a large-scale archive for testing visual devices and
  basic image processing algorithms.
\newblock In {\em STAG}, pages 63--70, 2014.

\bibitem{balle2018variational}
Johannes Ball{\'e}, David Minnen, Saurabh Singh, Sung~Jin Hwang, and Nick
  Johnston.
\newblock Variational image compression with a scale hyperprior.
\newblock {\em arXiv preprint arXiv:1802.01436}, 2018.

\bibitem{diffusioncompression2022}
Matthias B^^c3^^bchlmann.
\newblock Stable diffusion based image compression.
\newblock
  \url{https://pub.towardsai.net/stable-diffusion-based-image-compresssion-6f1f0a399202},
  2022.

\bibitem{chen2018encoder}
Liang-Chieh Chen, Yukun Zhu, George Papandreou, Florian Schroff, and Hartwig
  Adam.
\newblock Encoder-decoder with atrous separable convolution for semantic image
  segmentation.
\newblock In {\em Proceedings of the European conference on computer vision
  (ECCV)}, pages 801--818, 2018.

\bibitem{duan2023lossy}
Zhihao Duan, Ming Lu, Zhan Ma, and Fengqing Zhu.
\newblock Lossy image compression with quantized hierarchical vaes.
\newblock In {\em Proceedings of the IEEE/CVF Winter Conference on Applications
  of Computer Vision}, pages 198--207, 2023.

\bibitem{kodak2013}
R.~W Franzen.
\newblock True color kodak images.
\newblock \url{http://r0k.us/graphics/kodak/}, 2013.

\bibitem{goodfellow2014generative}
Ian Goodfellow, Jean Pouget-Abadie, Mehdi Mirza, Bing Xu, David Warde-Farley,
  Sherjil Ozair, Aaron Courville, and Yoshua Bengio.
\newblock Generative adversarial nets.
\newblock {\em Advances in neural information processing systems}, 27, 2014.

\bibitem{harvey2021conditional}
William Harvey, Saeid Naderiparizi, and Frank Wood.
\newblock Conditional image generation by conditioning variational
  auto-encoders.
\newblock {\em arXiv preprint arXiv:2102.12037}, 2021.

\bibitem{he2022elic}
Dailan He, Ziming Yang, Weikun Peng, Rui Ma, Hongwei Qin, and Yan Wang.
\newblock Elic: Efficient learned image compression with unevenly grouped
  space-channel contextual adaptive coding.
\newblock In {\em Proceedings of the IEEE/CVF Conference on Computer Vision and
  Pattern Recognition}, pages 5718--5727, 2022.

\bibitem{ho2020denoising}
Jonathan Ho, Ajay Jain, and Pieter Abbeel.
\newblock Denoising diffusion probabilistic models.
\newblock {\em Advances in neural information processing systems},
  33:6840--6851, 2020.

\bibitem{hoogeboom2021autoregressive}
Emiel Hoogeboom, Alexey~A Gritsenko, Jasmijn Bastings, Ben Poole, Rianne
  van~den Berg, and Tim Salimans.
\newblock Autoregressive diffusion models.
\newblock {\em arXiv preprint arXiv:2110.02037}, 2021.

\bibitem{isola2017image}
Phillip Isola, Jun-Yan Zhu, Tinghui Zhou, and Alexei~A Efros.
\newblock Image-to-image translation with conditional adversarial networks.
\newblock In {\em Proceedings of the IEEE conference on computer vision and
  pattern recognition}, pages 1125--1134, 2017.

\bibitem{mentzer2020high}
Fabian Mentzer, George~D Toderici, Michael Tschannen, and Eirikur Agustsson.
\newblock High-fidelity generative image compression.
\newblock {\em Advances in Neural Information Processing Systems},
  33:11913--11924, 2020.

\bibitem{morita2023batinet}
Ryugo Morita, Zhiqiang Zhang, and Jinjia Zhou.
\newblock Batinet: Background-aware text to image synthesis and manipulation
  network.
\newblock In {\em 2023 IEEE International Conference on Image Processing
  (ICIP)}, pages 765--769. IEEE, 2023.

\bibitem{ramesh2022hierarchical}
Aditya Ramesh, Prafulla Dhariwal, Alex Nichol, Casey Chu, and Mark Chen.
\newblock Hierarchical text-conditional image generation with clip latents.
\newblock {\em arXiv preprint arXiv:2204.06125}, 1(2):3, 2022.

\bibitem{richardson2021encoding}
Elad Richardson, Yuval Alaluf, Or~Patashnik, Yotam Nitzan, Yaniv Azar, Stav
  Shapiro, and Daniel Cohen-Or.
\newblock Encoding in style: a stylegan encoder for image-to-image translation.
\newblock In {\em Proceedings of the IEEE/CVF conference on computer vision and
  pattern recognition}, pages 2287--2296, 2021.

\bibitem{rombach2022high}
Robin Rombach, Andreas Blattmann, Dominik Lorenz, Patrick Esser, and Bj{\"o}rn
  Ommer.
\newblock High-resolution image synthesis with latent diffusion models.
\newblock In {\em Proceedings of the IEEE/CVF conference on computer vision and
  pattern recognition}, pages 10684--10695, 2022.

\bibitem{schuhmann2022laion}
Christoph Schuhmann, Romain Beaumont, Richard Vencu, Cade Gordon, Ross
  Wightman, Mehdi Cherti, Theo Coombes, Aarush Katta, Clayton Mullis, Mitchell
  Wortsman, et~al.
\newblock Laion-5b: An open large-scale dataset for training next generation
  image-text models.
\newblock {\em Advances in Neural Information Processing Systems},
  35:25278--25294, 2022.

\bibitem{taubman2002jpeg2000}
David~S Taubman and Michael~W Marcellin.
\newblock Jpeg2000: Standard for interactive imaging.
\newblock {\em Proceedings of the IEEE}, 90(8):1336--1357, 2002.

\bibitem{theis2022lossy}
Lucas Theis, Tim Salimans, Matthew~D Hoffman, and Fabian Mentzer.
\newblock Lossy compression with gaussian diffusion.
\newblock {\em arXiv preprint arXiv:2206.08889}, 2022.

\bibitem{wang2022neural}
Dezhao Wang, Wenhan Yang, Yueyu Hu, and Jiaying Liu.
\newblock Neural data-dependent transform for learned image compression.
\newblock In {\em Proceedings of the IEEE/CVF Conference on Computer Vision and
  Pattern Recognition}, pages 17379--17388, 2022.

\bibitem{yang2024lossy}
Ruihan Yang and Stephan Mandt.
\newblock Lossy image compression with conditional diffusion models.
\newblock {\em Advances in Neural Information Processing Systems}, 36, 2024.

\bibitem{yang2022diffusion}
Ruihan Yang, Prakhar Srivastava, and Stephan Mandt.
\newblock Diffusion probabilistic modeling for video generation, 2022.

\bibitem{zhang2023adding}
Lvmin Zhang, Anyi Rao, and Maneesh Agrawala.
\newblock Adding conditional control to text-to-image diffusion models.
\newblock In {\em Proceedings of the IEEE/CVF International Conference on
  Computer Vision}, pages 3836--3847, 2023.

\end{thebibliography}
\end{document}